\documentclass[preprint,tightenlines,aps,floats,nofootinbib]{revtex4}
\usepackage{epsf}
\usepackage{graphicx}

\newcommand{\lsim}
{\;\raisebox{-.3em}{$\stackrel{\displaystyle <}{\sim}$}\;}
\begin{document}

\title{Higgs Triplets, Decoupling, and Precision Measurements}
\author{Mu-Chun Chen$^{a}$}
\email[]{muchunc@uci.edu}
\author{Sally Dawson$^{b}$}
\email[]{dawson@bnl.gov}
\author{C.~B. Jackson${^c}$}
\email[]{jackson@hep.anl.gov}

\affiliation{
$^a$Department of Physics and Astronomy\\
University of California, 
Irvine,CA ~92697, USA \\
$^b$Department of Physics, Brookhaven National Laboratory, 
Upton, NY~ 11973, USA\\
$^c$ HEP Division, Argonne National Laboratory,9700 Cass Ave. Argonne, IL~60439 
\vspace*{.5in}}

\date{\today}

\begin{abstract}
Electroweak precision data has been extensively used to constrain models
containing physics beyond that of the Standard Model.  When the model
contains Higgs scalars in representations
other than $SU(2)$ singlets or doublets, and hence $\rho\ne 1$ at
tree level, a correct renormalization
scheme requires more inputs than the three needed for the Standard
Model.  We discuss the connection between the renormalization
of models with Higgs triplets and the decoupling
properties of the models as the mass scale
for the scalar triplet field becomes much larger than the electroweak
scale.
The 
requirements of perturbativity of the couplings and  agreement with 
electroweak data
place strong restrictions on models with Higgs triplets.  
Our results have important implications for Little Higgs
type models and other models with $\rho\ne 1$ at
tree level.   
\end{abstract}

\maketitle
\newpage

\section{Introduction}

The Standard Model of electroweak physics is remarkably successful at explaining
experimental data.  From a theoretical standpoint, however, the theory has many
failings.  Attempts to address these perceived inadequacies have led to the
construction of models which reduce to the Standard Model at energy
scales below about $1$ TeV, but which differ at higher energies. 
Models with physics beyond that of the Standard Model (SM), however,
 are severely
constrained by precision electroweak data\cite{Erler:2004nh,Erler:2008ek}.  
If the mass scale of
the new physics is near the TeV scale, it is often possible
to learn about the parameters of the model by performing
global fits to precision measurements. The simplest example
is the prediction of the $W$ boson mass, $M_W$. In the Standard Model, 
$M_W$ can be predicted in terms of other parameters
of the theory and requiring agreement with the measured $W$ mass therefore
restricts the possibilities for new $TeV$ scale physics.  

In this paper,
we introduce a Higgs triplet at the electroweak scale 
and consider the  effect on 
$M_W$\cite{Chen:2006pb,Chen:2005jx,Blank:1997qa,Lynn:1990zk}.
We are motivated by Little
Higgs models, which include a scalar triplet as a necessary 
ingredient,
although our results are very general\cite{ArkaniHamed:2001nc,ArkaniHamed:2002qy,ArkaniHamed:2002qx,Chang:2003un,Skiba:2003yf,Chang:2003zn,Schmaltz:2005ky,Chen:2006dy}.  
In a model with  Higgs particles in
representations other than $SU(2)$ doublets and singlets, there
are more parameters in the gauge/Higgs sector than in the Standard 
Model (SM). The SM tree level relation, $\rho=M_W^2/(M_Z^2 c_\theta^2)= 1$, 
no longer holds and when the theory is renormalized an extra input parameter
is required\cite{Lynn:1990zk,Passarino:1990xx,Passarino:1990nu,Czakon:1999ha,Czakon:1999ue,Chen:2006dy,Chen:2006pb,Chen:2005jx,Blank:1997qa}.
We discuss  two possible renormalization schemes for the triplet model:  
one where the extra
parameter is chosen to be a low energy 
observable\cite{Chen:2006pb,Chen:2005jx,Blank:1997qa}, and one where 
the extra parameter is taken to be the running vacuum
expectation value (VEV) of the triplet scalar\cite{Chankowski:2006hs,Chankowski:2007mf}.
Models with  $\rho=M_W^2/(M_Z^2 c_\theta^2)\ne 1$
can be consistent with experimental data with the inclusion of
certain types of new physics\cite{Peskin:2001rw,Chivukula:2000px}, 
of which a Higgs triplet is a 
specific example.

In Section \ref{themodel} we describe a model which contains a real
 Higgs
triplet in addition to the Higgs doublet of the Standard Model.
This example is a simplified version of the Higgs sector in Little
Higgs Models and is the simplest example of
a model with $\rho \ne 1$ at tree level.
  In Section \ref{limit}, we discuss the restrictions
on  models with scalar triplets at the electroweak scale
 from requiring
perturbativity of the parameters of the scalar 
potential.  
We turn in Section \ref{renorm} to a discussion of the renormalization
prescription and the role of the triplet vacuum expectation
value  (VEV).  The role of the scalar
particles is emphasized in obtaining predictions for $M_W$ in the
triplet model\cite{Hewett:2002px}.  
Whereas in the SM, the Higgs scalar contributes logarithmically
to 
the prediction for
$M_W$, in the triplet model there are contributions which grow with
the scalar 
masses- 
squared\cite{Toussaint:1978zm,Senjanovic:1978ee,Chen:2003fm,Forshaw:2001xq}.
We close in Section \ref{decoup} with a discussion
of the decoupling of Higgs triplet effects for large mass scales or
alternatively
in the limit that the triplet VEV goes to zero and we draw some general
conclusions about the renormalization scheme dependence in models
with $\rho \ne 1$ at tree limit.

\section{The Model}
\label{themodel}
We consider a model with a real 
Higgs doublet, $H$, and a real, isospin $Y=0$ triplet, $\Phi$. 
We assume that the scalar potential is such that the
neutral components of both the doublet and the triplet receive VEVs,
breaking the electroweak symmetry.  The
scalars are conventionally written as,
\begin{eqnarray}
H=\left(
\begin{array}{c}
\phi^+ \\
{1\over \sqrt{2}}(v+h^0+i\chi^0)
\end{array}
\right)\, ,
\qquad \qquad
\Phi=\left(
\begin{array}{c}
\eta^+\\
v^\prime+\eta^0\\
-\eta^-
\end{array}
\right)\, .
\end{eqnarray}
The kinetic part of the Lagrangian is
\begin{equation}
L=\mid D_\mu H\mid^2+{1\over 2}\mid D_\mu \Phi\mid^2 \, ,
\end{equation}
where,
\begin{eqnarray}
D_\mu H&=& \biggl( \partial_\mu +i {g\over 2}
 \sigma^aW^a + i{g^\prime\over 2} Y B_\mu\biggr)H
\nonumber \\
D_\mu\Phi&=&\biggl(\partial_\mu +i g t_aW^a\biggr)\Phi\, ,
\end{eqnarray}
$\sigma^a ~(a=1,2,3)$ are the Pauli matrices,
and
\begin{equation}
t_1={1\over \sqrt{2}}
\left(
\begin{array}{ccc}
0&1&0\\
1&0&1\\
0&1&0
\end{array}\right),\quad
t_2={1\over\sqrt{2}}
\left(\begin{array}{ccc}
0&-i&0\\
i&0&-i\\
0&i&0
\end{array}\right),\quad
t_3=\left(\begin{array}{ccc}
1&0&0\\
0&0&0\\
0&0&-1
\end{array}\right)\, .
\end{equation}
Spontaneous symmetry breaking generates masses for the $W$ and $Z$ bosons,
\begin{eqnarray}
M_W^2 &=& {g^2\over 4}\biggl( v^2+4 v^{\prime 2}\biggr)
\nonumber \\
M_Z^2 &=& {g^2\over 4 c_\theta^2}v^2\, ,
\end{eqnarray}
where $c_\theta\equiv \cos\theta=g/\sqrt{(g^\prime)^2+(g)^2}$ and $\sin\theta\equiv s_\theta$.
At tree level, all definitions of $c_\theta$ are equivalent
and the VEVs are related to the SM VEV by $v_{SM}^2=(246~GeV)^2=v^2+4 v^{\prime 2}$.

The  model violates custodial symmetry at tree level,
\begin{eqnarray}
\rho&=&{M_W^2\over M_Z^2 c_\theta^2}
=
1+4{v^{\prime 2}\over v^2}
\, .
\end{eqnarray}
Since experimentally $\rho\sim 1$,  $v^\prime$ will be 
restricted to be small.  Neglecting scalar loops, 
Refs. \cite{Erler:2004nh,Erler:2008ek} found $v^\prime
< 12~GeV$.

The most general $SU(2)\times U(1)$ scalar potential with a Higgs doublet and
a real scalar triplet is given by,
\begin{eqnarray}
V&=&\mu_1^2\mid H\mid^2
+{\mu_2^2\over 2}\mid \Phi\mid^2
+\lambda_1 \mid H\mid^4 
+{\lambda_2\over 4}\mid \Phi\mid^4
\nonumber \\ &&
+{\lambda_3\over 2}\mid H\mid^2\mid\Phi\mid^2
+\lambda_4  H^\dagger \sigma^a H\Phi_a \, .
\end{eqnarray}
We note that the coefficient $\lambda_4$ has dimensions of mass, which
implies that its effects may be important even for large mass scales
since the decoupling theorem is not applicable to interactions
which are proportional to mass\cite{Chen:2006pb,SekharChivukula:2007gi,Appelquist:1976wh}. 

After spontaneous symmetry breaking,
there are two  physical neutral eigenstates, $H^0,K^0$,
\begin{eqnarray}
\left(
\begin{array}{c}
H^0\\
K^0
\end{array}
\right)
=
\left(
\begin{array}{cc} 
c_\gamma & s_\gamma \\
-s_\gamma & c_\gamma 
\end{array}
\right)
\left(
\begin{array}{c}
h^0\\
\eta^0
\end{array}
\right)\, .
\end{eqnarray}
The physical charged eigenstates are $H^\pm$ and the
charged Goldstone bosons which become the longitudinal components of 
the $W^\pm$
bosons are $G^\pm$,
\begin{eqnarray}
\left(
\begin{array}{c}
G^+\\
H^+
\end{array}
\right)
=
\left(
\begin{array}{cc} 
c_\delta & s_\delta \\
-s_\delta & c_\delta 
\end{array}
\right)
\left(
\begin{array}{c}
\phi^+\\
\eta^+
\end{array}
\right)\, ,
\end{eqnarray}
where $\tan\delta =2 v^\prime/v$.

Minimizing the potential gives the tree-level relations:
\begin{eqnarray}
0&=&
\mu_1^2+\lambda_1 v^2+{\lambda_3\over 2}v^{\prime~2}-\lambda_4 v^\prime
\label{min1} \\
0&=&v^\prime\biggl(\mu_2^2 +\lambda_2 v^{\prime~2} +{\lambda_3}{v^2\over 2}
\biggr)-\lambda_4 {v^2\over 2}\, .
\label{min2}
\end{eqnarray}
For $v^\prime =0$, the only consistent solution to the minimization
of the potential has $M_{K^0}=M_{H^+}$ and $\lambda_4=\sin \delta=
\sin\gamma=0$.  In this limit the custodial symmetry is restored
and $\rho=1$ at tree level.

We assume that $v^\prime \ne 0$ and $\gamma\ne 0$.  
Utilizing Eqs. \ref{min1} and \ref{min2}, the scalar mass eigenstates 
are\cite{Forshaw:2003kh},
\begin{eqnarray}
M_{H^+}^2&=&{\lambda_4 v\over c_\delta s_\delta}
\nonumber \\
M_{H^0}^2&=&2 \lambda_1 v^2 +\tan \gamma (\lambda_3 v v^\prime -\lambda_4 v)
\nonumber \\
M_{K^0}^2&=& 2 \lambda_1 v^2 -\cot \gamma 
(\lambda_3 v v^\prime -\lambda_4 v) \, .
\label{mas}
\end{eqnarray}
We take as our $6$ input parameters in the scalar sector,
$M_{H^+},M_{H^0},M_{K^0},\gamma,\delta,v$.  
Inverting Eq. \ref{mas},
\begin{eqnarray}
\lambda_1&=& {1\over 2 v^2}\biggl(c_\gamma^2 M_{H^0}^2
+s_\gamma^2 M_{K^0}^2\biggr)
\nonumber \\
\lambda_2&=&{2\over v^2}\cot^2\delta\biggl[s_\gamma^2 M_{H^0}^2
+c_\gamma^2 M_{K^0}^2 -c_\delta^2M_{H^+}^2\biggr]
\nonumber \\
\lambda_3&=& {1\over v^2 \tan\delta}
\biggl[ 
(M_{H^0}^2-M_{K^0}^2)\sin (2 \gamma)
 +M_{H^+}^2 \sin (2 \delta)\biggr]
\nonumber \\
\lambda_4&=& c_\delta s_\delta {M_{H^+}^2\over v}
\nonumber \\
\mu_1^2&=&-{M_{H^0}^2\over 2}\biggl(c_\gamma^2+{s_\gamma c_\gamma\over 2}
\tan\delta\biggr)
-{M_{K^0}^2\over 2} \biggl( s_\gamma^2-{s_\gamma c_\gamma\over 2}\tan\delta
\biggr)+{M_{H^+}^2\over 4} s_\delta^2
\nonumber \\
\mu_2^2&=&{c_\delta^2\over 2} M_{H^+}^2
-{M_{H^0}^2\over 2}\biggl( s_\gamma^2+\sin(2\gamma)\cot\delta\biggr)
-{M_{K^0}^2\over 2}\biggl( c_\gamma^2-\sin(2\gamma)\cot\delta\biggr)\, .
\label{lamdefs}
\end{eqnarray}

\begin{figure}[t]
\begin{center}
\includegraphics[bb=14 19 351 279,scale=0.73]
{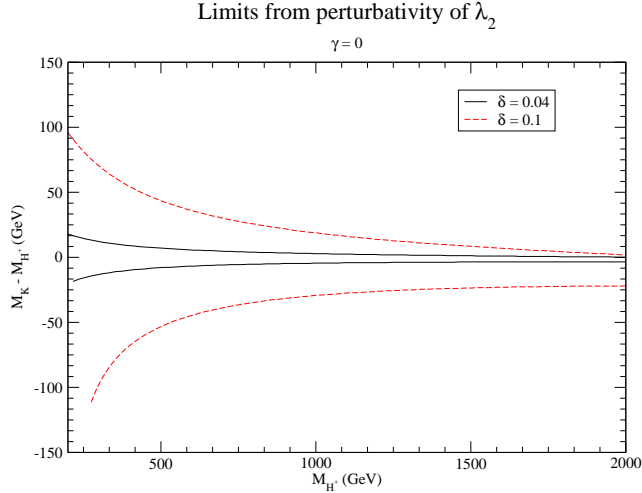}
\caption[]{Restriction on the mass difference, $M_{K^0}-M_{H^+}$ from the requirement
that the scalar coupling, $\lambda_2$, 
be perturbative, $\lambda_2\lsim (4\pi)^2$. For $\gamma=0$, there is
no dependence on $M_{H^0}$.}
\label{masslims} 
\end{center}
\end{figure}

\begin{figure}[t!]
\begin{center}
\includegraphics[bb=14 19 351 279,scale=0.73]
{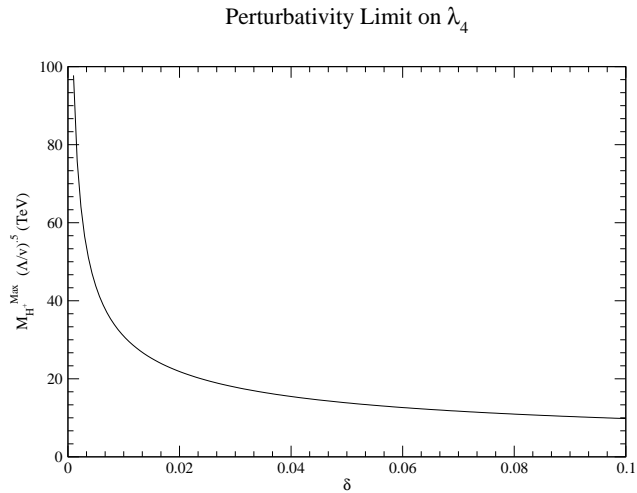}
\caption[]{Restriction on the charged scalar mass
from the requirement
that the scalar coupling, $\lambda_4$, 
be perturbative, ${\lambda_4\over \Lambda}\lsim (4\pi)^2$.}
\label{masslims2} 
\end{center}
\end{figure}

\section{Perturbativity of the Scalar Couplings}
\label{limit}

A priori, the input parameters, $M_{H^+},M_{H^0},M_{K^0},\gamma,\delta,v$,
are arbitrary.  
The requirement that the tree level contributions to the scalar self
interactions be larger than the one loop contributions can be
loosely interpreted as the restriction  $\lambda_{1,2,3} < (4\pi)^2$, and
$ {\lambda_4\over \Lambda} < (4\pi)^2$,  (the scale  $\Lambda$
is arbitrary.)

From Eq. \ref{lamdefs}, approximate bounds on the scalar masses can be 
derived\footnote{Ref. \cite{Forshaw:2003kh} derived the renormalization
group improved bounds on the scalar masses.  For our purposes the 
tree level bounds are sufficient.}.
The most restrictive bound on  $M_{H^0}$ is found
from $\lambda_1 \lsim (4\pi)^2$ for $c_\gamma\sim 1$,
\begin{equation}
M_{H^0} \lsim 4.4~TeV\, .
\end{equation}
An interesting limit on the mass difference between the charged scalar, $H^+$, and the
heavier of the neutral scalars, $K^0$, comes from Eq. \ref{lamdefs} and the requirement 
that $\lambda_2 \lsim (4\pi)^2$.  This restriction is illustrated 
in Fig. \ref{masslims}.
As the mass
of $M_{H^+}$ becomes large, perturbativity requires that the mass difference between
$M_{K^0}$ and $M_H^+$ be small, regardless of the mixing parameters.
Similarly, 
assuming a scale $\Lambda \sim v$, 
the perturbativity limits on $M_{H^+}$ from $\lambda_4$
are shown in Fig.\ref{masslims2}.  These results are in agreement
with those found in Refs. \cite{Forshaw:2003kh} and \cite{SekharChivukula:2007gi}.

\section{Renormalization}
\label{renorm}
\subsection{Standard Model}

In this section, we discuss the differences between renormalization in
the SM and in a model with a scalar triplet. 
We begin with a brief overview of 
Standard Model renormalization in order to set the 
framework\cite{Sirlin:1980nh,Sirlin:1981yz,Jegerlehner:1991dq,Hollik:1993xg}.
The electroweak sector of the SM has four fundamental 
parameters, the $SU(2)_{L} \times U(1)_{Y}$ gauge coupling constants, 
$g$ and $g^\prime$, the vacuum expectation value (VEV) of the 
neutral component of the Higgs doublet,  $v$,
 and the physical Higgs boson mass, along with the fermion masses.
Once these parameters are fixed, 
all other physical quantities in the gauge sector can be derived.
The usual choice of input parameters is the muon decay constant, $G_{\mu}$, 
the $Z$-boson mass, $M_{Z}$, the fine structure constant, 
$\alpha$, and the unknown Higgs boson mass, $M_{h,SM}$. 
Experimentally, the measured values for these input 
parameters are\cite{Amsler:2008zz},
\begin{eqnarray}
G_{\mu} & = & 1.16637(1) \times 10^{-5} \; GeV^{-2}\\
M_{Z} & = & 91.1876(21) \; GeV\\
\alpha^{-1} & = & 137.035999679(94) \quad .
\label{inputs}
\end{eqnarray}

Tree level objects are denoted with a subscript $0$ and satisfy the
relationship,
\begin{equation}
 M_{W0}^{2} = \frac{\pi \alpha_0}{\sqrt{2}G_{\mu 0 } s_{\theta 0}^{2}}
\, .
\end{equation}
and the SM satisfies $\rho_0 =1$ at tree level,
\begin{equation}
\rho_0 \equiv \frac{M_{W0}^{2}}{M_{Z0}^{2}c_{\theta 0}^{2}}=1\quad .
\label{rhodef}
\end{equation}

The $1$-loop renormalized quantitites are defined\footnote{Eq. \ref{dalp}
implicitely defines our sign conventions for $\Pi_{XY}$.  
We decompose the two-point functions as
$\Pi_{XY}^{\mu\nu}(k^2)= g^{\mu\nu}\Pi_{XY}(k^2)
+k^\mu k^\nu B_{XY}(k^2)$  and label the SM  contributions as $\Pi_{XY,SM}$.}
:
\begin{eqnarray}
M_{V0}^2&\equiv&
M_{V}^2\biggl(1+{\delta M_V^2\over M_V^2}\biggr)
=M_{V}^2\biggl(1+\Pi_{VV}(M_V^2)\biggr)
\nonumber \\
G_{\mu 0}&\equiv & G_\mu\biggl(1+{\delta G_\mu\over G_\mu}\biggr)
=G_\mu \biggl(1-{\Pi_{WW}(0)\over M_W^2}\biggr)
\nonumber \\
s_{\theta 0}^2&=&s_{\theta }^2 \biggl(1+{\delta s_\theta^2\over s_\theta^2}
\biggr)\nonumber \\
\alpha_0&=& \alpha\biggl(1+{\delta \alpha\over \alpha}\biggr)
=\alpha\biggl( 1+\Pi_{\gamma \gamma}^\prime (0)
+2{s_\theta\over c_\theta}
{\Pi_{\gamma Z}(0)\over M_Z^2} \biggr) \, ,
\label{dalp}
\end{eqnarray}
where $\Pi_{\gamma \gamma}^\prime (0)
=(d\Pi_{\gamma\gamma}(p^2)/dp^2)\mid_{p^2=0}
\sim  {\Pi_{\gamma \gamma}(M_Z^2)\over M_Z^2}$.
 The box and vertex corrections
are small and we neglect their finite contributions (although it is
necessary to include the poles in order to achieve a finite result).

The $W$-boson mass is predicted at $1$-loop,
\begin{equation}
 M_{W}^{2} = \frac{\pi \alpha}{\sqrt{2}G_{\mu} s_\theta^{2}}
\biggl[1 + \Delta r_{SM} \biggr]\, ,
\end{equation}
where $\Delta r_{SM}$ summarizes the radiative corrections,
\begin{eqnarray}
\Delta r_{SM} &=& - \frac{\delta G_{\mu}}{G_{\mu}}  
- \frac{\delta M_{W}^{2}}{M_{W}^{2}}
+\frac{\delta \alpha}{\alpha}
- \frac{\delta s_\theta^{2}}{s_\theta^{2}} 
\nonumber \\
&=&{\Pi_{WW,SM}(0)-\Pi_{WW,SM}(M_W^2)\over M_W^2}
+\Pi_{\gamma \gamma,SM}^\prime (0) +2{s_{\theta}\over c_{\theta}}
{\Pi_{\gamma Z,SM}(0)\over M_Z^2}
\nonumber \\ &&
-{\delta s^2_{\theta}\over s^2_{\theta}}
\quad .
\label{drdef}
\end{eqnarray}
We use $s_\theta\equiv\sin \theta,c_\theta\equiv \cos\theta$ 
to denote a generic definition of the weak mixing angle. At
tree level all definitions are equal and we consider three possible
definitions of the weak mixing angle, which differ
only at $1-$loop
and are useful for comparing with the predictions of the triplet model
in the next section and for
understanding the renormalization scheme dependence of the triplet model
predictions.  
For clarity, we review these definitions briefly\cite{Peskin:1995ev}.

\subsubsection{On-Shell Definition of $\sin \theta_W$}

In the on-shell scheme, the weak mixing angle, $s_W$,
 is not a free parameter, but is derived from
\begin{equation}
\rho\equiv {M_W^2\over M_Z^2 c_W^2}\quad .
\label{rhodefos}
\end{equation}
The counter term 
for $s_W^{2}$ can be derived from Eq. \ref{rhodefos}:
\begin{equation}
\frac{\delta s_W^{2}}{s_W^{2}}
=  \frac{c_W^{2}}{s_W^{2}} 
\biggl[ \frac{\delta M_{Z}^{2}}{M_{Z}^{2}} - \frac{\delta M_{W}^{2}}{M_{W}^{2}}
\biggr]
= \frac{c_W^{2}}{s_W^{2}}
\biggl[
\frac{\Pi_{ZZ,SM}(M_{Z}^{2})}{M_{Z}^{2}} - \frac{\Pi_{WW,SM}(M_{W}^{2})}{M_{W}^{2}}
\biggr]\quad \, .
\label{stdef}
\end{equation}

\subsubsection{Effective Mixing Angle Definition of $\sin^2\theta_{eff}$}

One could take as input parameters,
$G_\mu$, $\alpha$, and the effective weak mixing angle, $sin\theta_{eff}
\equiv s_{\theta,eff}$.
The effective weak mixing angle is defined in terms of the electron coupling
to the $Z$:
\begin{equation}
L = { e\over 2 \cos \theta_{eff} \sin\theta_{eff}}
{\overline e} \gamma_\mu\biggl( v_e -a_e \gamma_5\biggr) e Z^\mu
\end{equation}
where,
\begin{equation}
v_e=-{1\over 2} + 2 sin^2\theta_{eff} \qquad \qquad a_e=-{1\over 2}
\, .
\end{equation}
In this scheme, 
\begin{equation}
{\delta s_{\theta,eff}^2\over s_{\theta, eff}^2} =
\biggl({c_{\theta,eff}
\over s_{\theta,eff}}\biggr) 
      {\Pi_{\gamma Z,SM}(M_Z^2)\over M_Z^2}+{\cal O}(m_e^2)\, .
\label{seffdef}
\end{equation}
This scheme is useful for comparing with the predictions of the triplet
model using the renormalization 
scheme of Refs.\cite{Blank:1997qa,Chen:2005jx}.

\subsubsection{``$M_Z$'' Definition of $\sin \theta_Z$}
A third scheme for renormalizing the SM takes
as inputs $\alpha(M_Z), G_\mu$, and $M_Z$ and defines $sin \theta_Z
\equiv s_{Z}, \cos \theta_Z\equiv c_Z$ in terms of,
\begin{equation}
\sin (2\theta_Z)\equiv \sqrt{
{4 \pi \alpha(M_Z)\over \sqrt{2} G_\mu M_Z^2}}\, .
\label{drsm3}
\end{equation}
where
\begin{equation}
{\delta s_Z^2\over s_Z^2}=
{c_Z^2\over c_Z^2-s_Z^2}
\biggl(
{\Pi_{\gamma\gamma,SM}(M_Z^2)\over M_Z^2}
+{2 s_Z\over c_Z }{\Pi_{\gamma Z,SM}(0)
\over M_Z^2}
-{\Pi_{ZZ,SM}(M_Z^2)\over M_Z^2} +{\Pi_{WW,SM}(0)\over M_W^2}
\biggr) \, .
\end{equation}
The $s_Z$ scheme is useful for comparing with the predictions of the triplet
model using the renormalization 
scheme advocated in Ref. \cite{Chankowski:2006hs}.

\subsection{$Y=0$ Triplet}
In this section, we consider the $1-$loop renormalization of
the triplet model.  We are particularly interested in the approach
of the triplet model to the SM in different limits and in the scheme
dependence of our results.  
Since $\rho\ne 1$ at tree level in the triplet model, $4$ input parameters 
(along with the Higgs mass)
are required for the electroweak 
renormalization\cite{Blank:1997qa,Lynn:1990zk,Erler:2004nh,Chen:2006dy}. 
We will consider
two possible renormalization
schemes. The first scheme uses $4$ 
measured low energy parameters as inputs, while
the second employs $3$ low energy parameters plus a running 
triplet VEV, $v^\prime(\mu)$:
\begin{itemize}
\item
{\bf Scheme 1}: Input  $\alpha$, $M_Z$, $G_\mu$ and $\sin^2\theta_{eff}$
\item
{\bf Scheme 2}: Input
$\alpha$, $M_Z$, $G_\mu$ and $v^\prime (\mu)$.  
\end{itemize}
In both schemes, the $W$ boson mass is a predicted quantity.
Below we
discuss the dependence of the prediction 
for the $W$ mass on the renormalization scheme and
focus on the approach to the SM limit as the triplet VEV
becomes small, $v^\prime
\rightarrow 0$, or alternatively as $M_{K^0}$ and $M_{H^+}
\rightarrow \infty$.

\subsubsection{Triplet Model, Scheme 1}
The renormalization of the triplet model in Scheme 1 has been discussed
in Refs. \cite{Chen:2006pb,Chen:2005jx,Blank:1997qa}. The input 
parameters, $\alpha$, $M_Z$, and  $G_\mu$ are given in Eq. \ref{inputs},
and\cite{lepeww},
\begin{equation}
\sin^2\theta_{eff}=.2324\pm .0012\,\, .
\end{equation}
The relation,
\begin{equation}
\rho={1\over c_\delta^2}={M_W^2\over M_Z^2 c_{\theta,eff}^2}\, ,
\end{equation}
implies that $\tan \delta=2v^\prime/v$ is not a free parameter in this scheme,
but is fixed by the input parameters.

 In this scheme, 
 the $W$ mass is given by,
\begin{equation}
M_W^2={\alpha\pi\over\sqrt{2} s^2_{\theta, eff} G_\mu}
\biggl(1+\Delta r_{triplet}(Scheme~ 1)\biggr)
\label{mw1}
\end{equation}
and\footnote{We neglect the finite contribution from vertex and box
diagrams, although the pole contributions are included
in order to make our result finite and gauge invariant.
  The vertex and box corrections
 in the triplet model can be found in Ref.
\cite{Blank:1997qa}.}
\begin{eqnarray}
\Delta r_{triplet}(Scheme~ 1)&=&
{\Pi_{WW}(0)-\Pi_{WW}(M_W^2)\over M_W^2}
+\Pi_{\gamma \gamma}^\prime (0) +2{s_{\theta,eff}\over c_{\theta,eff}}
{\Pi_{\gamma Z}(0)\over M_Z^2}
-{\delta s^2_{\theta,eff}\over s^2_{\theta,eff}}
\label{drtriplet}
\end{eqnarray}
where
\begin{eqnarray}
{\delta s^2_{\theta,eff}
\over s^2_{\theta,eff}}&=&{c_{\theta,eff}\over s_{\theta,eff}}
{\Pi_{\gamma Z}(M_Z^2)\over M_Z^2}
\, .
\label{tripdefs}
\end{eqnarray}
Analytic formulae for the scalar, gauge boson, and Goldstone
boson  contributions to the  two-point functions 
are given in Appendix 1 for arbitrary values of the
mixing parameters $\sin \delta$ and $\sin\gamma$.  
The contributions from
non-zero values of $\sin \gamma$ have not appeared elsewhere.

There are three types of contributions to the two-point functions.  There
are contributions from the $W$,$Z$, and $\gamma$ gauge bosons, the
electroweak ghosts, the Goldstone bosons and the lightest neutral Higgs
boson where the couplings have SM strength.  These are labelled as $\Pi_{XY,SM}$ in Appendix 1.
It is important to remember that these are not numerically equal
to the results in the SM since the relationship between the $W$ and $Z$
masses is different in the SM and in the triplet model.  
The remaining contributions,
which we label $\Delta \Pi_{XY}$, are of two types.  There are contributions
from the SM particles with couplings proportional to $\sin\delta$ or
$\sin\gamma$ which vanish for $\delta, \gamma\rightarrow 0$, and there
are contributions from the new particles of the triplet model, $K^0$
and $H^+$, which do not necessarily vanish for $\delta, \gamma\rightarrow 0$.

Eq. \ref{drtriplet} has a form analogous to 
the SM results obtained using the $\sin\theta_{eff}$ scheme,
($\Delta r_{SM}^{eff}$),
except that in Eq. \ref{drtriplet} there are 
additional contributions to the two-point functions
 from $K^0$ and $H^+$, and the
SM-like gauge boson, Goldstone boson, 
and $H^0$ contributions are weighted by factors of $\cos\delta$
and $\cos\gamma$.
At tree level, the SM and triplet predictions for $M_W$ are
the same in this scheme.
It is useful to consider the difference between
  Eq. \ref{drtriplet} and the one-loop SM prediction,
\begin{equation}
\Delta r_{triplet}(Scheme~ 1)
=
 {\tilde{\Delta}} r^{eff}_{SM}+\Delta_{r,1}\, .
\end{equation}
The function ${\tilde{\Delta}} r^{eff}_{SM}$ has the same functional
form as $\Delta r^{eff}_{SM}$ with the important 
difference that in calculating ${\tilde{\Delta}} r^{eff}_{SM}$, $M_Z$
must be taken as an input in the triplet scheme 1, while $M_Z$ is a prediction
in the $\sin \theta_{eff}$ scheme of the SM. 
In the limit of small mixing ($\delta\sim 0, ~\gamma\sim 0$),
\begin{eqnarray}
\Delta_{r,1}&\rightarrow &
 {\alpha\over \pi \sin^2\theta_{eff}}\biggl[
{F_{22}(M_W^2,M_{K^0},M_{H^+})\over M_W^2}-
{F_{22}(M_Z^2,M_{H^+},M_{H^+})\over M_Z^2}\biggr]
\end{eqnarray}
where
\begin{equation}
F_{22}(k^2,m_1,m_2)=B_{22}(k^2,m_1,m_2)-B_{22}(0,m_1,m_2)
\, .
\end{equation}
The small mixing limit further simplifies in 
the limit that $M_{K^0},M_{H^+}>>M_W,M_Z$ and 
$\mid M_{K^0}^2-M_{H^+}^2\mid <<M_{K^0}^2$
\begin{equation}
 \Delta_{r,1} \rightarrow
{\alpha\over 24 \pi\sin^2\theta_{eff}}\biggl\{
{M_{K^0}^2-M_{H^+}^2\over M_{H^+}^2}
\biggr\}+....
\label{dr1dif}
\end{equation}
Eq. \ref{dr1dif} is independent of the light Higgs boson mass and
can be either positive or negative depending on the sign of $M_{K^0}-M_{H^+}$.

\begin{figure}[t]
\begin{center}
\includegraphics[angle=0,bb=20 25 335 276,scale=0.73]
{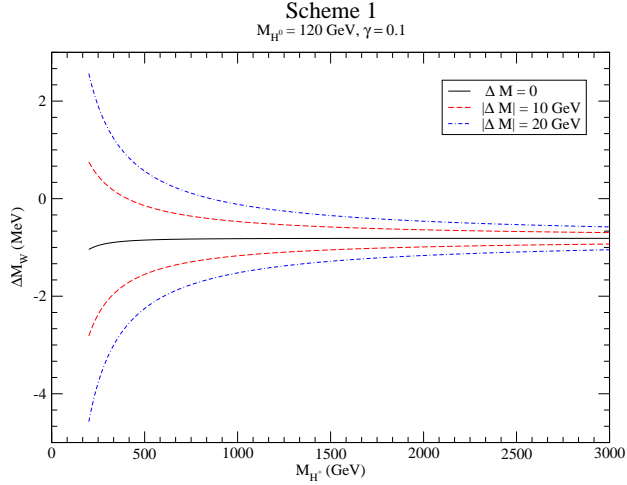}
\caption[]{Difference between the one-loop corrected values $M_W (Triplet,~Scheme~ 1)$ and
${M_W (SM,~ sin\theta_{eff}~Scheme })$ as a function of the charged Higgs
mass, $M_{H^+}$, for small 
mixing in the neutral sector, $\gamma=0.1$, 
(which corresponds to $v^\prime\sim 6.8~GeV$). The mass difference between the scalars is $\Delta M=M_{H^+}-M_{K^0}$.}
\label{dmws1} 
\end{center}
\end{figure}

In Fig. \ref{dmws1}, 
we show the approach of the triplet model
 to the one-loop SM prediction
 (in the $sin^2 \theta_{eff}$ renormalization scheme)
as $M_{H^+}$ becomes large.    
The SM prediction for $M_W$ is calculated using Eqs. \ref{drdef} and 
\ref{seffdef}, while
the triplet prediction for $M_W$ is calculated using Eqs. 
\ref{drtriplet} and \ref{tripdefs}.
For small mixing, and  $M_{K^0}= M_{H^+}$, the one loop prediction for
$M_W$ in the triplet model
differs negligibly from the SM prediction.  As the mass splitting,
$\mid M_{K^0}-M_{H^+}\mid$ 
is increased, significant differences from the SM prediction are
seen at small $M_{H^+}$.
The remainder term, $\Delta_{r,1}$, never
goes exactly to zero, because the triplet model has $M_Z$ as an
input, while the SM computes $M_Z$
in the $sin\theta_{eff}$ scheme.  
We recall from Section \ref{themodel} that in 
the limit $M_{K^0}=M_{H^+}$, the only consistent
solution to the minimization of the potential is $v^\prime=0$
and $ c_\delta
=c_\gamma=1$.  
In this limit, $\rho_0=1$ and the only difference between
the prediction of triplet model and the SM arises from the
different input values of $M_Z$.

\subsubsection{Triplet Model, Scheme 2}

In Scheme 2, the input parameters are $\alpha,M_Z, G_\mu$ and 
a running $v^\prime(\mu)$.
This scheme has been advocated in Ref. \cite{Chankowski:2006hs} as being
more natural than Scheme 1, in that it has $3$ measured input  parameters
as does the SM, while $v^\prime$ is unknown. We will treat $v^\prime$ as
a running ${\overline {MS}}$ parameter. Of particular interest
is the $v^\prime \rightarrow 0$ limit and the approach to the SM
as $M_{K^0}$ and $M_{H^+}\rightarrow \infty$.

As usual,  the $W$ boson mass is defined by,
\begin{equation}
M_W^2={\alpha\pi\over\sqrt{2} s_{\theta}^2 G_\mu}
\biggl(1+\Delta r_{triplet}(Scheme~ 2)\biggr)
\label{mw2}
\end{equation}
where\footnote{Again, we neglect the small finite contributions from
the vertex and box diagrams, although the pole contributions are included
in order to make our result finite and gauge independent.}
\begin{eqnarray}
\Delta r_{triplet}(Scheme~ 2)&=&
{\Pi_{WW}(0)-\Pi_{WW}(M_W^2)\over M_W^2}
+{\Pi_{\gamma \gamma}(M_Z^2)\over M_Z^2} 
+{2 {\hat s}_Z\over
{\hat c}_Z}{\Pi_{\gamma Z}(0)\over M_Z^2}
\nonumber \\ &&
-{\delta {\hat s}_Z^2 \over {\hat s}_Z^2}
\, .
\label{drtriplet2}
\end{eqnarray}
At tree level,
\begin{equation}
G_{\mu 0}={1\over \sqrt{2}(v_0^2+4 v_0^{\prime 2})}\, ,
\label{gdef}
\end{equation}
where $v_0$ and $v_0^\prime$ are the tree level VEVs.
Using
\begin{equation}
M_{Z_0}^2={e_0^2\over 4 {\hat s}_{Z 0}^2 
{\hat c}_{Z 0}^2 } v_0^2
\end{equation}
and Eq. \ref{gdef},
we find the weak mixing angle,
\begin{eqnarray}
{\hat s}_{Z 0}^2 {\hat c}_{Z 0}^2
={\pi\alpha_0 v_0^2 \over M_{Z_0}^2}
={\pi \alpha_0\over M_{Z_0}^2 }\biggl[{1\over \sqrt{2} G_{\mu 0}}-4 v_0^{\prime 2}
\biggr]\, .
\label{snew}
\end{eqnarray}
This scheme is similar  to the $M_Z$ scheme for the
SM described in Eq. \ref{drsm3} in the limit $v^\prime_0 \rightarrow 0$. 
At tree level ${\hat s}_{Z 0}$ 
is defined in terms of the input
parameters as
\begin{equation}
{\hat s}_{Z 0}^2 ={1\over 2}\biggl( 1-\sqrt{1-{4 \pi \alpha_0\over \sqrt{2}
M_{Z_0}^2 G_{\mu 0}}(1-4\sqrt{2} G_{\mu 0} v_0^{\prime 2})}\biggr) \, ,
\label{relation}
\end{equation}
and the $1-$loop corrected value for the mixing angle is 
\begin{eqnarray}
{\delta {\hat s}_Z^2\over {\hat s}_Z^2} &=&
{{\hat c}_Z^2\over {\hat c}_Z^2-{\hat s}_Z^2}
\biggl\{
\Pi^\prime_{\gamma\gamma}(0)
+{2{\hat s}_Z\over {\hat c}_Z}
{\Pi_{\gamma Z}(0)\over M_Z^2}
-{\Pi_{ZZ}(M_Z^2)\over M_Z^2} 
\nonumber \\ &&
+
{1\over 1-4 \sqrt{2} v^{\prime 2} G_{\mu}}
\biggl[
{\Pi_{WW}(0)\over M_W^2}-4\sqrt{2} G_\mu v^{\prime 2}
{\delta v^{\prime 2}\over v^{\prime 2}}\biggr]
\biggr\} \, .
\label{ds2}
\end{eqnarray}
Finally, we need to define 
the renormalized triplet vev: 
\begin{equation}
v_0^{\prime 2}=v^{\prime 2}(1+{\delta
v^{\prime 2}\over v^{\prime 2}})\, . 
\label{vpren}
\end{equation}
There is no compelling physical definition for the $v^\prime$
  counterterm and
we simply utilize an ${\overline{MS}}$ definition and retain only the 
 poles necessary to cancel the divergences.
\begin{figure}[t]
\begin{center}
\includegraphics[angle=0,bb=20 25 335 276,scale=0.73]
{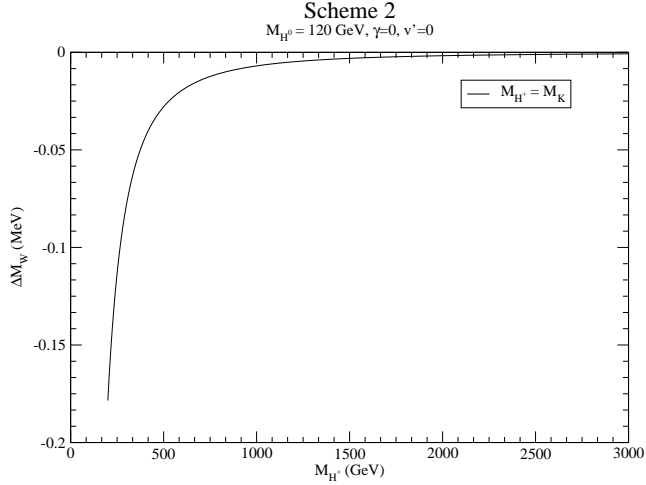}
\caption[]{Difference between the one-loop corrected values $M_W (Triplet,~Scheme~2 )$ and
${M_W (SM,~ M_Z~Scheme})$ as a function of the charged Higgs
mass, $M_{H^+}$, for zero 
mixing in the neutral sector ($\gamma=0$) and for $v^\prime=0$.}
\label{dmws2_fig} 
\end{center}
\end{figure}

In Fig. \ref{dmws2_fig}, we compare the one-loop corrected prediction for $M_W$ in the $M_Z$
scheme of the SM, with the one-loop corrected value for $M_W$ in the triplet model, Scheme 2,
with $\gamma=0$ and $v^\prime=0$.  For $\gamma=0$ and $v^\prime=0$, the only consistent
solution to the minimization of the potential is $M_{H^+}=M_{K^0}$ and for these parameters,
the contribution of the triplet model quickly decouples as $M_{H^+}$ becomes large and
the SM result is exactly recovered.  

\begin{figure}[t]
\begin{center}
\includegraphics[angle=0,bb=20 25 335 276,scale=0.73]
{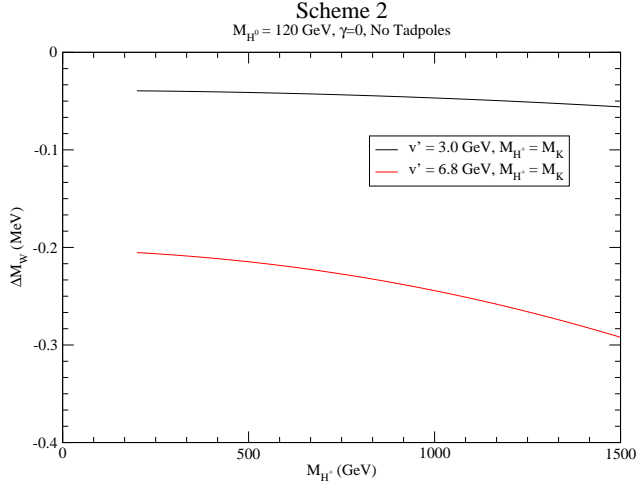}
\caption[]{Difference between the one-loop corrected values 
$M_W (Triplet,~Scheme~2 )$ and
$M_W (SM,~ M_Z~Scheme)$ as a function of the charged Higgs
mass, $M_{H^+}$, for no
mixing in the neutral sector ($\gamma=0.0$) and for $v^\prime=3~GeV$ and
$v^\prime=6.8~GeV$. Tadpoles are not included in this plot.}
\label{schem2a} 
\end{center}
\end{figure}

\begin{figure}[t]
\begin{center}
\includegraphics[angle=0,bb=20 25 335 276,scale=0.73]
{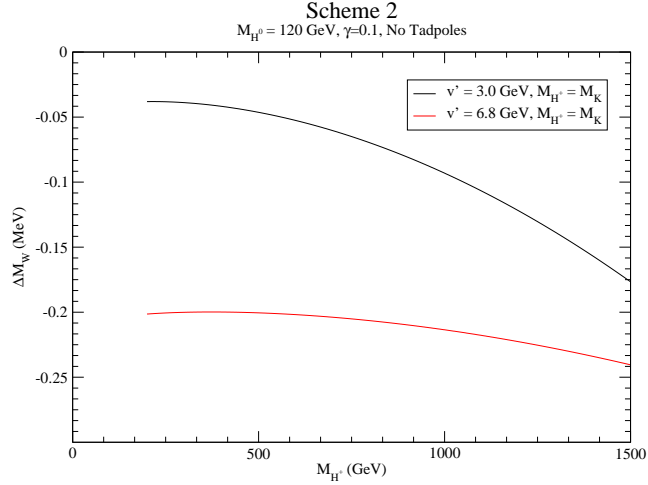}
\caption[]{Difference between the one-loop corrected values 
$M_W (Triplet,~Scheme~2 )$ and
$M_W (SM,~ M_Z~Scheme)$ as a function of the charged Higgs
mass, $M_{H^+}$, for small
mixing in the neutral sector ($\gamma=0.1$) and for $v^\prime=3~GeV$ and
$v^\prime=6.8~GeV$. Tadpoles are not included in this plot.}
\label{schem2b} 
\end{center}
\end{figure}

The
situation is quite different for non-zero $v^\prime$ as shown in 
Figs. \ref{schem2a} and \ref{schem2b}.
The large effects can be understood from Eq. \ref{ds2},
\begin{eqnarray}
{\delta {\hat s}_Z^2\over {\hat s}_Z^2} &\sim &
{{\hat c}_Z^2\over {\hat c}_Z^2-{\hat s}_Z^2}
\biggl\{
-{\Pi_{ZZ}(M_Z^2)\over M_Z^2} 
+
{1\over 1-4 \sqrt{2} v^{\prime 2} G_{\mu}}
{\Pi_{WW}(0)\over M_W^2}
\biggr\}\, .
\label{ds3}
\end{eqnarray}
Both ${\Pi_{ZZ}(M_Z^2)\over M_Z^2}$ and 
${\Pi_{WW}(0)\over M_W^2}$ have contributions which grow with $M_{H^+}^2$ and
$M_{K^0}^2$ which cancel in Eq. \ref{ds3} when $v^\prime=0$.  The cancellation
is spoiled for non-zero $v^\prime$ leading to the large effects observed
in Figs. \ref{schem2a} and \ref{schem2b}.  Figs. 
\ref{schem2a} and \ref{schem2b} show a modestsensitivity of our results
to non-zero mixing in the neutral sector ($\gamma\ne 0$).

\begin{figure}[t]
\begin{center}
\hskip 1.5in
\includegraphics[angle=0,bb=169 674 264 766, scale=.9]
{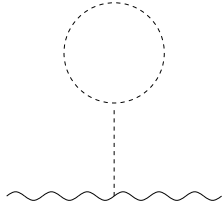}
\vskip 1in
\caption[]{Tadpole contribution to the gauge boson two-point function.}
\label{fg:tadpole} 
\end{center}
\end{figure}

\begin{figure}[t]
\begin{center}
\includegraphics[angle=0,bb=20 25 335 276,scale=0.73]
{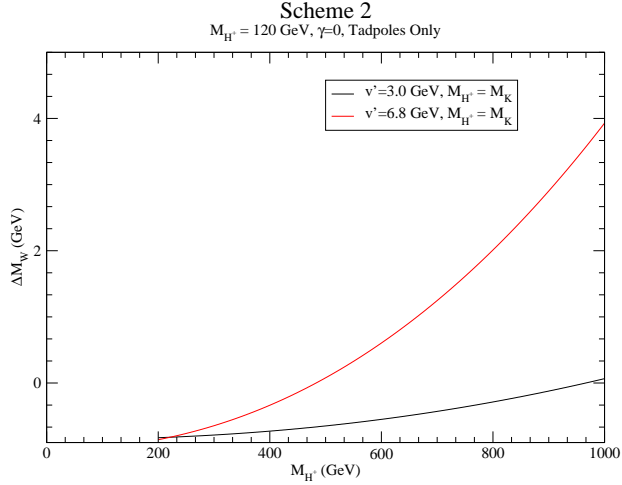}
\caption[]{Difference between the tadpole contribution to $M_W$ 
in Scheme 2 
and the one-loop prediction for
$M_W (SM,~ M_Z~Scheme)$ as a function of the charged Higgs
mass, $M_{H^+}$ for zero
mixing in the neutral sector ($\gamma=0.0$) and for 
$v^\prime=5~GeV$ and $v^\prime=9~GeV$.}
\label{tadsa} 
\end{center}
\end{figure}

\begin{figure}[t]
\begin{center}
\includegraphics[angle=0,bb=20 25 335 276,scale=0.73]
{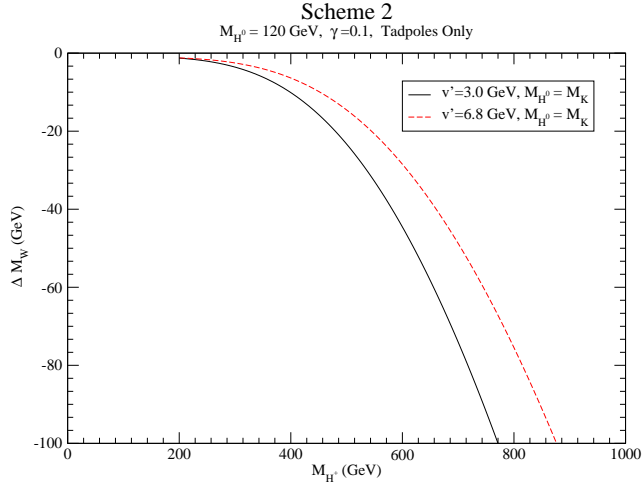}
\caption[]{Difference between the tadpole contribution to $M_W$ 
in Scheme 2 
and the one-loop prediction for
$M_W (SM,~ M_Z~Scheme)$ as a function of the charged Higgs
mass, $M_{H^+}$ for small
mixing in the neutral sector ($\gamma=0.1$) and for 
$v^\prime=5~GeV$ and $v^\prime=9~GeV$.}
\label{tadsb} 
\end{center}
\end{figure}

Eq. \ref{ds3} makes it  apparent that tadpole diagrams 
(shown in Fig. \ref{fg:tadpole})
do not cancel for non-zero $v^\prime$ and make a contribution,
\begin{equation}
\Delta r_{triplet}(Scheme~2)^{tadpole}=-
{{\hat c}_Z^2\over {\hat c}_Z^2-{\hat s}_Z^2}
\biggl\{
-{\Pi_{ZZ}^{tadpole}\over M_Z^2}
+{1\over 1-4\sqrt{2} v^{\prime 2}G_\mu}
{\Pi_{WW}^{tadpole}\over M_W^2}\biggr\}\, ,
\end{equation}
where the tadpole contributions are,
\begin{eqnarray}
\Pi_{WW}^{tadpole}&=& -gM_W\biggl\{
(c_\delta c_\gamma+2 s_\delta s_\gamma){{\cal T_H}\over M_{H^0}^2}
+
(-c_\delta s_\gamma+2 s_\delta c_\gamma){{\cal T_K}\over M_{K^0}^2}
\biggr\}
\nonumber \\
\Pi_{ZZ}^{tadpole}&=& -g{M_Z\over {\hat c}_Z}
\biggl\{ c_\gamma{{\cal T_H}\over M_{H^0}^2}
-
s_\gamma {{\cal T_K}\over M_{K^0}^2}
\biggr\}\, .
\end{eqnarray}
The scalar self couplings 
are given in Appendix 3\cite{Chankowski:2007mf} and lead to,
\begin{eqnarray}
{\cal T}_H&=& -{1\over 16\pi^2}\biggl\{ 
{g_{HHH}\over 2}A_0(M_{H^0})
+{g_{HKK}\over 2}A_0(M_{K^0})
+{g_{HG^0G^0}\over 2}A_0(M_Z)
+g_{HG^+G^-}A_0(M_W)\nonumber \\
&&+g_{H^0H^+H^-}A_0(M_{H^+})
-gM_W(c_\delta c_\gamma+2 s_\delta s_\gamma)(4-2\epsilon)A_0(M_W)
\nonumber \\ &&
-gM_Z{c_\gamma\over c_\theta}(4-2\epsilon)A_0(M_Z)\biggr\}
\nonumber \\
{\cal T}_K&=& -{1\over 16\pi^2}\biggl\{ 
{g_{KHH}\over 2}A_0(M_{H^0})
+{g_{KKK}\over 2}A_0(M_{K^0})
+{g_{KG^0G^0}\over 2}A_0(M_Z)
+g_{KG^+G^-}A_0(M_W)\nonumber \\
&&+g_{K^0H^+H^-}A_0(M_{H^+})
-gM_W(-c_\delta s_\gamma+2 s_\delta c_\gamma)(4-2\epsilon)A_0(M_W)
\nonumber \\ &&
+gM_Z{s_\gamma\over c_\theta}(4-2\epsilon)A_0(M_Z)\biggr\} \, .
\end{eqnarray}
The tadpole diagrams generate terms which grow with mass-squared and the contribution
of the tadpole diagrams in Scheme 2 to $M_W$
 are shown in Figs. \ref{tadsa} and \ref{tadsb}. The large size of
the tadpole contributions makes it clear that 
$\delta v^{\prime}$ must be defined in such a manner as to cancel the contributions
from the tadpole diagrams in order to have a sensible theory. The tadpole
contributions grow with $v^{\prime 2}$ as expected and have a large
dependence on $\gamma$.

\section{Decoupling and Conclusions}
\label{decoup}
We have considered the simplest possible model with $\rho\ne 1$ at tree level:
a model with a real scalar $SU(2)$ triplet in addition to the SM Higgs doublet
and have presented results for the one-loop prediction
for the $W$ mass in two different renormalization schemes. Our results are shown
as differences from the SM predictions.
A correct renormalization scheme  in the triplet model 
requires four input parameters, in contrast
to the three required in the electroweak sector of the SM.

In the first scheme, four low energy measured parameters are used as inputs and
the theory is renormalized as a low energy theory.  The effects of the scalar
loops are negligible for large triplet scalar masses, when the mass difference
between the 
scalar masses associated
with the triplets  is
small ($\mid M_{K^0} -M_{H^+}\mid << M_{K^0}$).  This renormalization scheme fixes
the triplet $v^\prime$ in terms of the input parameters and so the limit $v^\prime
\rightarrow 0$ cannot be taken.
In the second scheme, three low energy parameters and a running triplet VEV
are used as inputs.  The non-zero triplet VEV generates large contributions
from tadpole diagrams which must be cancelled by hand by an appropriate 
definition of the triplet VEV renormalization condition.  This fine tuning 
implies a lack of predictivity for the model.  Neither renormalization scheme
is entirely satisfactory, although our results clearly demonstrate the 
importance of scalar loops in theories with $\rho \ne 1$ at tree level.

\section*{Acknowledgements}
The work of S.D. (C.J.) is supported by the U.S. Department of Energy under 
grant DE-AC02-98CH10886 (DE-AC02-06CH11357). The work of M.-C. C. is supported, in part,
by the National Science Foundation under grant PHY-0709742.
 S.D. thanks the SLAC theory group for their hospitality,
where this work was begun.

\section*{Appendix 1: 2-Point contributions}

In general, we write
\begin{equation}
\Pi_{XY}(k^2)=\Pi_{XY,SM}(k^2)+\Delta \Pi_{XY}(k^2)\, .
\end{equation}
The contributions labelled $\Pi_{XY,SM}$ have the same functional form
as the SM contributions
 from gauge and Goldstone bosons, ghosts, and the lightest neutral
Higgs, $H^0$, in the $\rho_0=1$ limit.  We remind the reader yet again 
that the    $\Pi_{XY,SM}$ terms utilize different relations between $M_Z$ 
and $M_W$ in the triplet and SM and hence are not in general numerically
equal.
The remainder, $\Delta \Pi_{XY}$, contains terms
which vanish in the limit $\sin\delta, \sin\gamma\rightarrow 0$, and
also the contributions of $K^0$ and $H^{+}$, which need not vanish in
the $\sin\delta=\sin\gamma=0$ limit. 
The SM-like contributions agree with those found in Ref. \cite{Hollik:1993xg} and the
triplet contributions for $\sin\gamma=0$ agree with Ref. \cite{Blank:1997qa}. The
contributions for non-zero $\gamma$ are new. 
\begin{figure}[tb!]
\begin{center}
\includegraphics{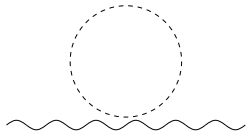}  
\end{center}
\caption[]{Contributions from Goldstone Bosons and $H^0,K^0$
and $H^+$ to the gauge boson two-point functions. }
\label{fg:wwss}
\end{figure}

From Fig. \ref{fg:wwss}, the Standard Model-like contributions in Feynman
gauge are:
\begin{eqnarray}
\Pi_{WW,SM}^1(k^2)&=& {\alpha\over 16\pi s_\theta^2}\biggl\{
A_0(M_{H^0}) + A_0(M_Z)
+2 A_0(M_W)\biggr\}\nonumber \\
\Pi_{ZZ,SM}^1(k^2)&=& {\alpha\over 16\pi s_\theta^2 c_\theta^2}
\biggl\{A_0(M_{H^0}) +A_0(M_Z)
+2(c_\theta^2-s_\theta^2)^2 A_0(M_W)\biggr\}
\nonumber \\
\Pi_{\gamma\gamma,SM}^1(k^2)&=& 
{\alpha\over 2 \pi} 
\biggl[ A_0(M_W)\biggr]
\nonumber \\
\Pi_{\gamma Z,SM}^1(k^2)&=& 
{\alpha\over 4 \pi s_\theta c_\theta}
\biggl[
(s_\theta^2-c_\theta^2)
A_0(M_W)\biggr]\, .
\end{eqnarray}
The non-Standard Model contributions from  Fig. \ref{fg:wwss} are:
\begin{eqnarray}
\Delta \Pi_{WW}^1(k^2)&=& 
{\alpha\over 16\pi s_\theta^2}\biggl\{-3 s_\gamma^2
\biggl[A_0(M_{K^0})- A_0(M_{H^0}\biggr]
+2s_\delta^2\biggl[A_0(M_W)- A_0(M_{H^+})\biggr]
\nonumber \\ && +4 \biggl[ A_0(M_{K^0})+A_0(M_{H^+})\biggr]\biggr\}
\nonumber \\
\Delta \Pi_{ZZ}^1(k^2)&=& {\alpha\over 16\pi s_\theta^2 c_\theta^2}
\biggl\{
s_\gamma^2\biggl[ A_0(M_{K^0})-A_0(M_{H^0})\biggr]
 +2s_\delta^2 (1-4 c_\theta^2)\biggl[A_0(M_{H^+})
-A_0(M_W)\biggr] \nonumber \\
&&
+8 c_\theta^4 A_0(M_{H^+})\biggr\}
\nonumber \\
\Delta \Pi_{\gamma\gamma}^1(k^2)&=& 
{\alpha\over 2 \pi} 
A_0(M_{H^\pm})
\nonumber \\
\Delta \Pi_{\gamma Z}^1(k^2)&=& 
{\alpha\over 4 \pi s_\theta c_\theta}
\biggl\{
s_\delta^2\biggl[
A_0(M_{H^+})-A_0(M_W)\biggr]
-2 c_\theta^2 A_0(M_{H^+})\biggr\}\, .
\end{eqnarray}

\begin{figure}[t]
\begin{center}
\includegraphics[bb=104 611 349 704,scale=0.63]
{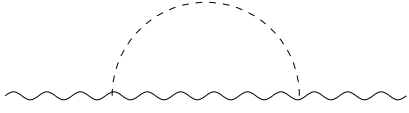}  
\end{center}
\caption[]{Contribution to the gauge boson 2-point function with one internal scalar
and one internal vector boson.}
\label{fg:wws}
\end{figure}

From Fig. \ref{fg:wws}, the SM-like contributions are,
\begin{eqnarray}
\Pi_{WW,SM}^2(k^2)&=&{\alpha\over 4 \pi}{M_W^2\over s_\theta^2}
\biggl\{
{s_\theta^4\over c_\theta^2}B_0(k^2,M_Z,M_W)
+s_\theta^2 B_0(k^2,0,M_W)
+B_0(k^2,M_{H^0},M_W)\biggr]
\biggr\}\nonumber \\
\Pi_{ZZ,SM}^2(k^2)&=&{\alpha\over 4 \pi}{M_Z^2\over s_\theta^2}
\biggl\{
{1\over c_\theta^2} B_0(k^2,M_{H^0}, M_Z)
+2 s_\theta^4\
 B_0(k^2,M_W,M_W)\biggr\}
\nonumber \\
\Pi_{\gamma\gamma,SM}^2(k^2)&=&{\alpha\over 2 \pi} M_W^2 B_0(k^2, M_W,M_W)
\nonumber \\ 
\Pi_{\gamma Z,SM}^2(k^2)&=& {\alpha\over 2 \pi} M_W^2 {s_\theta
\over  c_\theta}B_0(k^2, M_W, M_W)\, .
\end{eqnarray}
The non-Standard Model contributions from Fig. \ref{fg:wws} are:
\begin{eqnarray}
\Delta \Pi_{WW}^2(k^2)&=&{\alpha\over 4 \pi}{M_W^2\over s_\theta^2}
\biggl\{
{c_\delta^2s_\delta^2\over c_\theta^2}
\biggl[
B_0(k^2,M_Z, M_{H^+})
-B_0(k^2,M_Z,M_W)
\biggr]
\nonumber \\
&&
+\biggl(
4 s_\delta c_\delta s_\gamma c_\gamma-s_\gamma^2+5s_\delta^2 s_\gamma^2
\biggr)
\biggl[
B_0(k^2,M_{H^0},M_W)-
B_0(k^2,M_{K^0},M_W)
\biggr]
\nonumber \\
&&
+s_\delta^2
\biggl[
{c_\theta^2-s_\theta^2\over c_\theta^2} B_0(k^2,M_Z,M_W)
-B_0(k^2,M_{H^0},M_W)
+4 B_0(k^2,M_{K^0},M_W)\biggr]
\biggr\}
\nonumber \\
\Delta \Pi_{ZZ}^2(k^2)&=&
{\alpha\over 4 \pi}{M_Z^2\over s_\theta^2}
\biggl\{
{s_\gamma^2\over c_\theta^2}\biggl[
 B_0(k^2,M_{K^0},M_Z)-
 B_0(k^2,M_{H^0}, M_Z)\biggr]
\nonumber \\
&&
+2s_\delta^2
\biggl[
 B_0(k^2,M_{H^+},M_W)-
 B_0(k^2,M_W, M_W)\biggr]
\nonumber \\
&&+2{s_\delta^2\over c_\delta^2}c_\theta^4
 B_0(k^2,M_W,M_W)
\biggr\}
\nonumber \\
\Delta\Pi_{\gamma\gamma}^2(k^2)&=& 0
\nonumber \\ 
\Delta \Pi_{\gamma Z}^2(k^2)&=& -{\alpha\over 2 \pi} M_W^2 {
s_\delta^2\over s_\theta c_\theta}B_0(k^2, M_W, M_W) \, .
\end{eqnarray}

\begin{figure}[tb!]
\begin{center}
\includegraphics{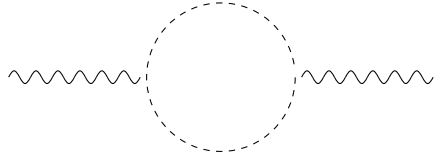}  
\end{center}
\caption[]{Contribution to the gauge boson 2-point 
function from 
Goldstone boson and scalar loops.}
\label{fg:wss}
\end{figure}
From Fig. \ref{fg:wss}, the SM-like contributions are
\begin{eqnarray}
\Pi_{WW,SM}^3(k^2)&=&-{\alpha\over 4 \pi s_\theta^2}\biggl\{
B_{22}(k^2, M_{H^0}, M_{W})
+ B_{22}(k^2, M_Z, M_W)\biggr\}\nonumber \\
\Pi_{ZZ,SM}^3(k^2)&=& -{\alpha\over 4 \pi s_\theta^2 c_\theta^2}
\biggl\{ B_{22}(k^2,M_{H^0},M_Z)
+(c_\theta^2-s_\theta^2)^2
B_{22}(k^2,M_W,M_W)
\biggr\}\nonumber \\
\Pi_{\gamma\gamma,SM}^3(k^2)&=& -{\alpha\over \pi}
B_{22}(k^2,M_W,M_W)\nonumber \\
\Pi_{\gamma Z,SM}^3(k^2)&=& {\alpha\over 2 \pi c_\theta s_\theta}
(c_\theta^2-s_\theta^2)B_{22}(k^2, M_W,M_W)\, .
\end{eqnarray}
From Fig. \ref{fg:wss}, the non-SM contributions are,
\begin{eqnarray}
\Delta \Pi_{WW}^3(k^2)&=&-{\alpha\over 4 \pi s_\theta^2}
\biggl\{
4 s_\gamma c_\gamma s_\delta c_\delta  
\biggr( 
B_{22}(k^2, M_{K^0}, M_{H^+})-
B_{22}(k^2, M_{H^0}, M_{H^+})
\nonumber \\
&&
+B_{22}(k^2, M_{H^0}, M_{W})
-B_{22}(k^2, M_{K^0}, M_{W})
\biggr)
\nonumber \\
&& +4s _\gamma^2  s_\delta^2
\biggl(
B_{22}(k^2, M_{H^0}, M_{W})
-B_{22}(k^2,M_{H^0},M_{H^+})
\biggr)
\nonumber \\
&&
+s_\gamma^2 s_\delta^2
\biggl(
B_{22}(k^2, M_{K^0}, M_{H^+})-
B_{22}(k^2,M_{H^0},M_{H^+})
\biggr)
\nonumber \\
&&
+4 c_\gamma^2 s_\delta^2 
\biggl(
 B_{22}(k^2, M_{K^0}, M_{W})-
B_{22}(k^2, M_{K^0}, M_{H^+})
\biggr)
\nonumber \\
&&
+s_\gamma^2 c_\delta^2
\biggl(
 B_{22}(k^2, M_{K^0}, M_{W})
- B_{22}(k^2, M_{H^0}, M_{W})
\biggr)
\nonumber \\
&&
+s_\delta^2 
\biggl(
 B_{22}(k^2, M_Z, M_{H^+})- B_{22}(k^2, M_Z, M_{W})
\nonumber \\
&&
+ B_{22}(k^2, M_{H^0},M_{H^+})
- B_{22}(k^2, M_{H^0}, M_{W})
\biggr)
\nonumber \\
&&
+4 s_\gamma^2\biggl(
B_{22}(k^2, M_{H^0},M_{H^+})-
B_{22}(k^2, M_{K^0},M_{H^+})
\biggr)
\nonumber \\
&&
+4B_{22}(k^2,M_{K^0},M_{H^+})\biggr\}
\nonumber \\
\Delta \Pi_{ZZ}^3(k^2)&=& -{\alpha\over 4 \pi s_\theta^2 c_\theta^2}
\biggl\{ 
s_\gamma^2\biggl[ B_{22}(k^2, M_{K^0},M_Z)- B_{22}(k^2,M_{H^0},M_Z)\biggr]
\nonumber \\
&&+s_\delta^2\biggl[ 4 c_\theta^2
\biggl(B_{22}(k^2,M_W,M_W)-
B_{22}(k^2, M_{H^+},M_{H^+})\biggr)
\nonumber \\ &&
+s_\delta^2
\biggl(
B_{22}(k^2, M_{H^+}, M_{H^+})-
B_{22}(k^2, M_{H^+}, M_W)
\nonumber \\ &&
+ B_{22}(k^2,M_W,M_W)-
 B_{22}(k^2, M_{H^+}, M_W)\biggr)
\nonumber \\ &&
+2\biggl( B_{22}(k^2, M_{H^+}, M_W)-B_{22}(k^2,M_W,M_W)\biggr)
\biggr]
\nonumber \\ && + 4 c_\theta^4 B_{22}(k^2, M_{H^+},M_{H^+})
\biggr\}\nonumber \\
\Delta \Pi_{\gamma\gamma}^3(k^2)&=&- 
{\alpha\over \pi} B_{22}(k^2, M_{H^+},
M_{H^+})\nonumber \\
\Delta \Pi_{\gamma Z}^3(k^2)&=& {\alpha\over 2 \pi c_\theta s_\theta}\biggl\{
s_\delta^2\biggl[
B_{22}(k^2,M_W,M_W)-B_{22}(k^2, M_{H^+},M_{H^+})\biggr]
\nonumber \\ &&  + 2 c_\theta^2B_{22}(k^2, M_{H^+},M_{H^+})
\biggr\}\, .
\end{eqnarray}
There are additional contributions which only contribute to $\Pi_{SM}$, which
we list for completeness\cite{Hollik:1993xg}.
\begin{figure}[tb!]
\begin{center}
\hskip -.65in
\includegraphics[bb=109 593 272 661,scale=0.63]
{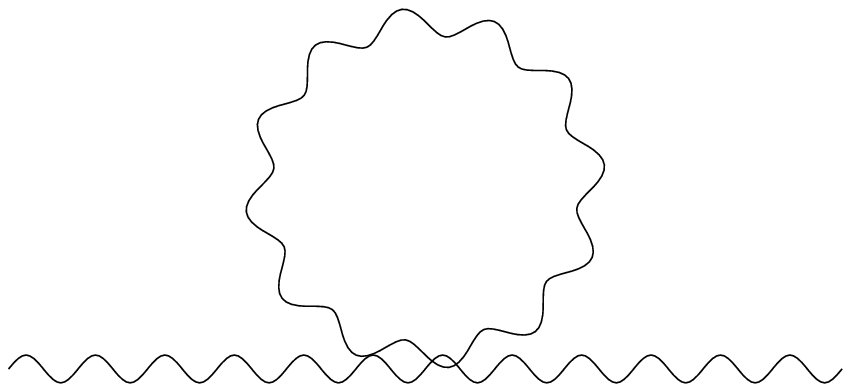}  
\end{center}
\caption[]{Contribution to gauge boson two-point function from SM gauge bosons in loop.}
\label{fg:pi4pt}
\end{figure}
From Fig. \ref{fg:pi4pt},
\begin{eqnarray}
\Pi_{WW,SM}^4(k^2)&=&{\alpha\over 4 \pi s_\theta^2}(3-2\epsilon) 
\biggl[ A_0(M_W)+c_\theta^2 A_0(M_Z)\biggr] \nonumber \\
\Pi_{ZZ,SM}^4(k^2)&=&{\alpha c_\theta^2 \over 2 \pi s_\theta^2}
(3-2\epsilon) 
A_0(M_W) \nonumber \\
\Pi_{\gamma\gamma,SM}^4(k^2)&=&{\alpha\over 2 \pi} 
(3-2\epsilon) 
A_0(M_W) \nonumber \\
\Pi_{\gamma Z,SM}^4(k^2)&=&- {\alpha c_\theta \over 2 \pi s_\theta}
(3-2\epsilon) 
A_0(M_W)\, .
\end{eqnarray}

\begin{figure}[t]
\begin{center}
\includegraphics[bb=109 593 272 661,scale=0.83]
{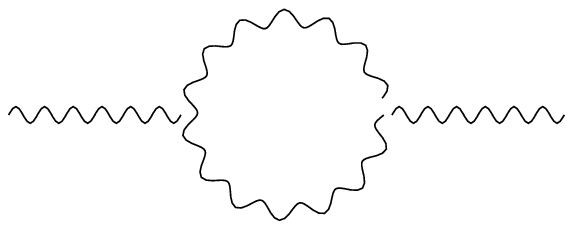}   
\end{center}
\caption[]{Contribution to the gauge boson two-point function from
 SM gauge bosons in loop.}
\label{fg:pi5pt}
\end{figure}
From Fig. \ref{fg:pi5pt},
\begin{eqnarray}
\Pi_{WW,SM}^5(k^2)&=&{\alpha\over 4 \pi s_\theta^2} 
\biggl[s_\theta^2 A_1(k^2,0,M_W)+c_\theta^2 A_1(k^2,M_Z,M_W)\biggr] \nonumber \\
\Pi_{ZZ,SM}^5(k^2)&=&{\alpha c_\theta^2 \over 4 \pi s_\theta^2}
A_1(k^2,M_W,M_W) \nonumber \\
\Pi_{\gamma\gamma,SM}^5(k^2)&=&{\alpha\over 4 \pi} 
A_1(k^2,M_W,M_W) \nonumber \\
\Pi_{\gamma Z,SM}^5(k^2)&=& -{\alpha c_\theta \over 4 \pi s_\theta}
A_1(k^2,M_W,M_W)\, ,
\end{eqnarray}
where,
\begin{eqnarray}
A_1(k^2,m_1,m_2)&=& -A_0(m_1)-A_0(m_2)-\biggl(m_1^2+m_2^2+4k^2\biggr)
B_0(k^2,m_1,m_1)
\nonumber \\
&&-10B_{22}(k^2,m_1,m_2)+2\biggl(m_1^2+m_2^2-{k^2\over 3}\biggr)\, .
\end{eqnarray}

\begin{figure}[t]
\vskip 1 in
\begin{center}
\vskip 1 in
\includegraphics{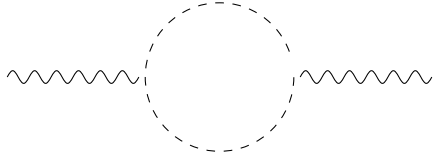}  
\end{center}
\caption[]{Contribution to gauge boson two-point function from ghosts in loop.}
\label{fg:pi6pt}
\end{figure}

From Fig. \ref{fg:pi6pt},
\begin{eqnarray}
\Pi_{WW,SM}^6(k^2)&=&{\alpha\over 2 \pi s_\theta^2} 
\biggl[s_\theta^2 B_{22}(k^2,0,M_W)
+c_\theta^2 B_{22}(k^2,M_Z,M_W)\biggr] \nonumber \\
\Pi_{ZZ,SM}^6(k^2)&=&{\alpha c_\theta^2 \over 2 \pi s_\theta^2}
B_{22}(k^2,M_W,M_W) \nonumber \\
\Pi_{\gamma\gamma,SM}^6(k^2)&=&{\alpha\over 2 \pi} 
B_{22}(k^2,M_W,M_W) \nonumber \\
\Pi_{\gamma Z,SM}^6(k^2)&=& -{\alpha c_\theta \over 2 \pi s_\theta}
B_{22}(k^2,M_W,M_W)\, .
\end{eqnarray}

\section*{Appendix 2:  Scalar Integrals}
The scalar integrals in $n=4-2\epsilon$ dimensions are defined as,
\begin{eqnarray}
{i\over 16\pi^2}A_0(m)&=&\int {d^n q\over (2\pi)^n}{1\over q^2-m^2}
\nonumber \\
{i\over 16\pi^2}B_0(k^2,m_1,m_2)&=&\int {d^n q\over (2\pi)^n}
{1\over [q^2-m_1^2][(q+k)^2-m_2^2]} \, .
\end{eqnarray}
The tensor integral is,
\begin{equation}
{i\over 16\pi^2}\biggl\{
g^{\mu\nu}B_{22}(k^2,m_1,m_2)
+k^\mu k^\nu B_{12}(k^2,m_1,m_2)\biggr\}
=\int {d^n q\over (2\pi)^n}
{q^\mu q^\nu\over [q^2-m_1^2][(q+k)^2-m_2^2]}.
\end{equation}

\section*{Appendix 3:  Scalar Self-Couplings}

\begin{eqnarray}
g_{HHH}&=&6\biggl(v c_\gamma^3 \lambda_1+v^\prime s_\gamma^3 \lambda_2
+{c_\gamma s_\gamma\over 2}(s_\gamma v+c_\gamma v^\prime)\lambda_3
-{s_\gamma c_\gamma^2\over 2} \lambda_4\biggr)
\nonumber \\
g_{HHK}&=&2\biggl(-3v s_\gamma c_\gamma^2 \lambda_1
+3 s_\gamma^2 c_\gamma v^\prime \lambda_2+{\lambda_3\over 2}\biggl[
-vs_\gamma (1-3 c_\gamma^2)+v^\prime c_\gamma(1-3s_\gamma^2)\biggr]
\nonumber \\&& \qquad 
-{c_\gamma\over 2}(1-3s_\gamma^2)\lambda_4 
\biggr)\nonumber \\
g_{HKK}&=&2\biggl(3v\lambda_1 c_\gamma s_\gamma^2
+3v^\prime \lambda_2c_\gamma^2 s_\gamma
+{\lambda_3\over 2}\biggl[
vc_\gamma (1-3 s_\gamma^2)+v^\prime s_\gamma(1-3c_\gamma^2)\biggr]
\nonumber \\&& \qquad 
-{s_\gamma\over 2}(1-3c_\gamma^2)\lambda_4 
\biggr)\nonumber \\
g_{KKK}&=&6\biggl(- \lambda_1 v s_\gamma^3+\lambda_2 v^\prime c_\gamma^3
+{c_\gamma s_\gamma\over 2}(-c_\gamma v+s_\gamma v^\prime)\lambda_3 
-{c_\gamma s_\gamma^2\over 2}\lambda_4 
\biggr)
\nonumber \\
g_{ HG^+G^-}&=&\biggl(
2 v\lambda_1 c_\gamma c_\delta^2
+2v^\prime \lambda_2 s_\gamma s_\delta^2
+\lambda_3(v c_\gamma s_\delta^2+v^\prime s_\gamma c_\delta^2)
\nonumber \\ && \qquad
+c_\delta \lambda_4(2 c_\gamma s_\delta +s_\gamma c_\delta)
\biggr)
\nonumber \\
g_{H H^+H^-}&=&\biggl(
2 v\lambda_1 c_\gamma s_\delta^2
+2v^\prime \lambda_2 s_\gamma c_\delta^2
+\lambda_3(v c_\gamma c_\delta^2+v^\prime s_\gamma s_\delta^2)
\nonumber \\ && \qquad
-s_\delta \lambda_4(2 c_\gamma c_\delta -s_\gamma s_\delta)
\biggr)
\nonumber \\
g_{K G^+G^-}&=&\biggl(
-2 v\lambda_1 s_\gamma c_\delta^2
+2v^\prime \lambda_2 c_\gamma s_\delta^2
+\lambda_3(-v s_\gamma s_\delta^2+v^\prime c_\gamma c_\delta^2)
\nonumber \\ && \qquad
+c_\delta \lambda_4(-2 s_\gamma s_\delta +c_\gamma c_\delta)
\biggr)
\nonumber \\
g_{K H^+H^-}&=&\biggl(
-2 v\lambda_1 s_\gamma s_\delta^2
+2 v^\prime \lambda_2 c_\gamma c_\delta^2
+\lambda_3 (-v s_\gamma c_\delta^2 +v^\prime c_\gamma s_\delta^2)
\nonumber \\ && \qquad
+s_\delta \lambda_4(2 s_\gamma c_\delta+ c_\gamma s_\delta)
\biggr)
\nonumber \\
g_{HG^0G^0}&=& \biggl(\lambda_1 c_\gamma v+{\lambda_3\over 2}v^\prime 
s_\gamma-{\lambda_4\over 2}c_\gamma\biggr)
\nonumber \\
g_{KG^0G^0}&=& \biggl(-\lambda_1 s_\gamma v+{\lambda_3\over 2}v^\prime 
c_\gamma+{\lambda_4\over 2}s_\gamma\biggr)
\end{eqnarray}

\bibliography{triplet}

\end{document}